# Engineering the electronic structure of TiO$_2$ by transition metal doping: A First Principles DFT Study

Vikash Mishra[1], Shashi Pandey[2], Swaroop Ganguly[2], Alok Shukla[3*]

[1]Department of Physics, Manipal Institute of Technology, Manipal Academy of Higher Education, Manipal, 576104, Karnataka, India

[2]Department of Electrical Engineering Indian Institute of Technology Bombay, Powai, Mumbai 400076, India

[3]Department of Physics, Indian Institute of Technology Bombay, Powai, Mumbai 400076, India

## Abstract

By means of first-principles density-functional theory (DFT) calculations, we perform a comparative analysis of the electronic and magnetic properties of transition metal doped TiO$_2$. The electronic band gaps of Ti$_x$M$_{1-x}$O$_2$, where M represents 3d-transition metals such as Sc, V, Cr, Mn, Fe, Co, Ni, Cu, and Zn, have been determined using the PBE functional within the generalized-gradient approximation (GGA) scheme, and hybrid HSE06 functional. In the context of pure TiO$_2$, the partial density of states (PDOS) reveals that the electronic band gap emerges between the O-2p and Ti-3d orbitals. It is suggested that the Ti-3d ($t_{2g}$) states play a more prominent role in bonding compared to the Ti-3d ($e_g$) states. We performed DFT calculations to investigate the impact of doping with other 3d transition metal atoms leading to the emergence of impurity states within the band gap. The hybridization between the oxygen 2p orbitals and the titanium 3d orbitals in TiO$_2$ is altered by the introduction of doping with 3d transition metals because of the change in the oxidation state of titanium, shifting from solely 4+ to a combination of 4+ and 3+ states. The calculation of spin-polarized density demonstrates the emergence of ferromagnetic properties, particularly in titanium dioxide doped with chromium (Cr), manganese (Mn), and iron (Fe) with large magnetic moments. Our work demonstrates the significant impact of doping transition metals on TiO$_2$, allowing for the precise manipulation of electrical and magnetic properties, and thus holds great potential for the development of spin-based memory devices with possible neuromorphic applications.

# Introduction

Wide band-gap transition-metal oxides, have attracted extensive research interest due to their structural flexibility, chemical stability, and diverse range of applications in photocatalysis, sensors, and transparent electronics [1-7]. In particular, $TiO_2$, due to its tunable electronic and magnetic properties, is increasingly investigated for its potential to develop exciting properties such as localized defect states, modulation of carrier concentration, and even magnetic ordering through doping with 3d-transition-metal atoms [8-11]. In doped $TiO_2$, these effects have led to observations of room-temperature ferromagnetism, resistive switching behavior, and enhanced photocatalytic activity, making it a versatile platform for both fundamental studies and device-oriented research. Despite these advances, the microscopic mechanisms driving these emergent behaviors remain poorly understood, underscoring the need for systematic experimental and theoretical investigations. For instance, recent studies have demonstrated the significance of titanium dioxide ($TiO_2$), or its doping with 3d-transition metal atoms, in elucidating the electronic characteristics in the vicinity of the Fermi level [12,13,14]. The investigation of changes in the magnetic characteristics of $TiO_2$ when doped with transition metals is important. Numerous theoretical studies have demonstrated that the introduction of transition metals into $TiO_2$ results in the emergence of a magnetic moment in the vicinity of the dopant atom, as well as in its nearest neighboring atoms. Similarly, Extensive experimental studies on 3d-transition-metal doped $TiO_2$ (i.e. $Co^{2+}$ and $Fe^{3+}$ dopants) have reported emergent phenomena including room-temperature ferromagnetism, enhanced conductivity, and improved photocatalytic activity. In Co-doped anatase $TiO_2$, strong magneto-optical effects and saturation magnetizations up to ~1 $\mu_B$/Co have been observed [8,9]; however, detailed spectroscopic (i.e. XPS, ESR) and microscopic studies reveal that ferromagnetism often arises from the intrinsic point defects (e.g., $Ti^{3+}$ or Ti vacancies) or from Co clustering rather than uniform substitutional doping. Oxygen-deficient processing conditions (e.g., vacuum annealing) tend to enhance ferromagnetic ordering by promoting bound magnetic polarons or vacancy-mediated exchange, whereas oxidizing conditions suppress it due to the formation of nonmagnetic phases. Similarly, Fe-doped and Fe–Cu co-doped $TiO_2$ powders exhibit robust room-temperature ferromagnetism arising from overlapping magnetic polarons and vacancy defects [10, 11]. Consequently, the investigation of the characterization of these materials in the solid-state phase has become a subject of interest for various research groups. However, thus far, the majority of research has focused on experimental approaches, with limited attention given to first principle calculations [15-18]. Several previous studies have examined the electronic band structure and electronic density of states of pure metal oxides [8,14]. However, the investigation of comparative impurity effects in wide band gap semiconductors remains unexplored. The existence of three prevalent crystal forms, namely Anatase, Rutile, and Brookite, is widely acknowledged in the scientific community for the compound $TiO_2$.

A study carried out in 2002; Park et al. [13] achieved successful growth of ferromagnetic Co-doped rutile $TiO_2$ films [19-21]. The Curie temperature ($T_C$) was calculated to be above 400 K for Co with a doping level of 12%. The present study investigates the electrical and magnetic characteristics of transition metal-doped $TiO_2$, specifically $Ti_{1-x}M_xO_2$, where M represents a range of 3d-transition metals including Sc, V, Cr, Mn, Fe, Co, Ni, Cu, and Zn. In photocatalytic applications, anatase is frequently chosen over rutile due to its higher band gap, which results in a longer charge-carrier lifetime and increased photocatalytic activity [18]. Keeping this in view we have focused our current investigation on the anatase phase of $TiO_2$ [22]. The presence of magnetism in Fe-doped rutile $TiO_2$ film has been revealed by experimental investigation [20]. Similarly, the introduction of Cu-doping in rutile $TiO_2$ also results in a significant magnetic moment, which can be attributed to the existence of oxygen vacancies [21,23]. However, as far as current knowledge is concerned, there has been no comparative investigation conducted on the electronic structure and magnetism of 3d-transition metal doped anatase $TiO_2$. The phenomenon of magnetism in transition-metal doped $TiO_2$ can be comprehended through spin polarization resulting from alterations caused by the dopant atom. This leads to the promotion of an electron from the $t_{2g}$ orbitals to the $e_g$ orbitals, which can induce varying modifications in the surrounding environment. The lower transition metals exhibit significant spin-orbit coupling, leading to the occurrence of Pseudo Jahn-Teller distortions [24-26] in octahedral structures. These distortions result in changes in spin polarization, which have been found in the $O_1$-Ti-$O_2$ bonds as well as the $O_1$-M-$O_2$ bonds in the $Ti_{1-x}M_xO_2$ compound. The enhancement of magnetic moments inside the system is a consequence of the alteration in spin polarization. Conversely, it is commonly acknowledged that the dependence of TM d-d transitions [27,28] on the hybridization of the doping orbitals with the host element is also recognized. The d-d transition occurring between the $t_{2g}$ and $e_g$ orbitals in the O1-Ti-O2-M-O1-Ti-O2 structure of metal-doped $TiO_2$ results in modifications in spin polarity, which varies depending on the dopant atom (M) incorporated into the TiO$_2$ lattice. The band gap of transition metal-doped TiO$_2$ is not directly associated with the band gap between the Ti t$_{2g}$ (d$_{xy}$, d$_{xz}$, d$_{yz}$) and e$_g$ (d$z^2$, d$x^2-y^2$) bands. Instead, it is determined by the energy separation between the overlapping of O-2p and the Ti-3d bands of TiO$_2$. Consequently, this characteristic is also influenced by the presence of doping atoms.

In this present work we have performed a comparative study on the magnetic and electronic properties of 3d-transition metal doped anatase TiO$_2$ using a density-functional theory (DFT) based first principles approach. We have performed calculations using various approximations, however, results obtained from the HSE06 calculations were found to be very close to the experimental values. For the 3d-transition metal doped structures Ti$_x$M$_{1-x}$O$_2$ (M=Sc, V Cr, Mn, Fe, Co, Ni, Cu, Zn)[14], the HSE06 calculations were performed the using 2x2x1 super cells. The DFT band structure calculations on the doped structures reveal the presence of impurity bands in the midgap region caused by the hybridization between O-2p, Ti-3d, and M-3d orbitals causing a change in oxidation state of Ti from 4+ to mixed states (i.e., 4+ and 3+). The spin-polarized projected DOS calculations confirm ferromagnetic behavior for most of the compounds, while the maximum magnetic moments can be seen especially in Cr, Mn and Fe doped TiO$_2$.

**Methods and Computational Details:**

Density-functional theory was employed to perform spin-polarized calculations for the total energy as well as the band structure and projected density of states (PDOS). These calculations utilized projected-augmented wave (PAW) potentials and were carried out using the Quantum Espresso (QE) simulation software [29–34]. The results of the HSE06 calculations were found to be very close to the experimental values, despite the fact that we had also done preliminary calculations using various approximations, including LDA and the generalized gradient approximation (GGA) utilizing the PBE functional to verify the bandgap values. The calculations were performed on the super cell consisting of two-unit cells arranged in a stacked manner along the a-axes. For the anatase phase of TiO$_2$, the super cell consists of 48 atoms in all, with 16 Ti atom and 32 O atoms (Ti$_{16}$O$_{32}$), in which one titanium (Ti) atom is substituted with a 3d-transition metal atom (denoted as M) selected among the atoms: Scandium (Sc), Vanadium (V), Chromium (Cr), Manganese (Mn), Iron (Fe), Cobalt (Co), Nickel (Ni), Copper (Cu), and Zinc (Zn). The crystal structures of the anatase phase of TiO$_2$ and the 3d-doped TiO$_2$ are depicted in figure 1, where the Ti atoms are represented by blue spheres, the O atoms by green spheres, and the transition metal dopant atoms by red spheres. The convergence of the total energy with respect to the size of the k-mesh is shown in figure 2(a) based on which we used 7×7×1 k-mesh and for density of states calculations we have 13×13×1 k-mesh. The super cells of the doped structures have the chemical formula $T_{15}M_1O_{32}$ (M= Sc, V, Cr, Mn, Fe, Co, Ni, Cu, Zn), amounting to a doping concentration of 6.25%. Both the pure and doped $TiO_2$ structures are fully relaxed until the forces/atom approach 0.04 eV/Å and an energy convergence of $5\times10^{-5}$ eV is reached in order to guarantee correct results. A 400 eV kinetic energy threshold [35,36] was used in our computations.

**Results and Discussion:**

Prior to the calculations of electronic and magnetic properties of the systems under consideration, we optimized the geometry of each structure. From Figure 2 (b) it is clearly seen that due to doping the optimized lattice are different in each case. Experimentally, the band gap of titanium dioxide has been estimated around 3.22 eV from various reports with the indirect band gap nature[4,37]. To match with experimental bandgap we have tried various approximations such as LDA, GGA using the PBE functional, and the hybrid functional HSE06 [30]. LDA is the most basic DFT functional, based on the idea that the electron density at a particular location is the only factor influencing the exchange-correlation energy there [17]. In general, it underestimates band gaps and can result in inaccurate lattice parameters and other characteristics, despite its computational efficiency. PBE is a generalized gradient approximation (GGA) functional that incorporates the electron density gradient into the exchange-correlation energy computation, thereby improving upon LDA [12]. PBE tends to underestimate band gaps but generally outperforms LDA for structural features like cohesive energies and lattice constants. It's a widely used functional due to its balance between accuracy and computational cost. The hybrid functional HSE06 introduces non-local exchange interactions by combining Hartree-Fock exchange with a screened Coulomb potential. HSE06 is a widely used option for investigating electronic properties because of this non-local exchange, which is essential for accurately estimating band gaps [31]. Compared to LDA or PBE, HSE06 calculations require more computing power but frequently yield noticeably more precise band gaps [1,18,30,38,39]. The band gap of $TiO_2$ computed using various approximations has been tabulated in table-1, and it is found that using the HSE06 functional, it is nearly ~ 3.2 eV which is extremely close to the experimental value 3.22 eV[4,37]. Here no extra electron correlation term has been added, i.e., no U has been used in our calculations.

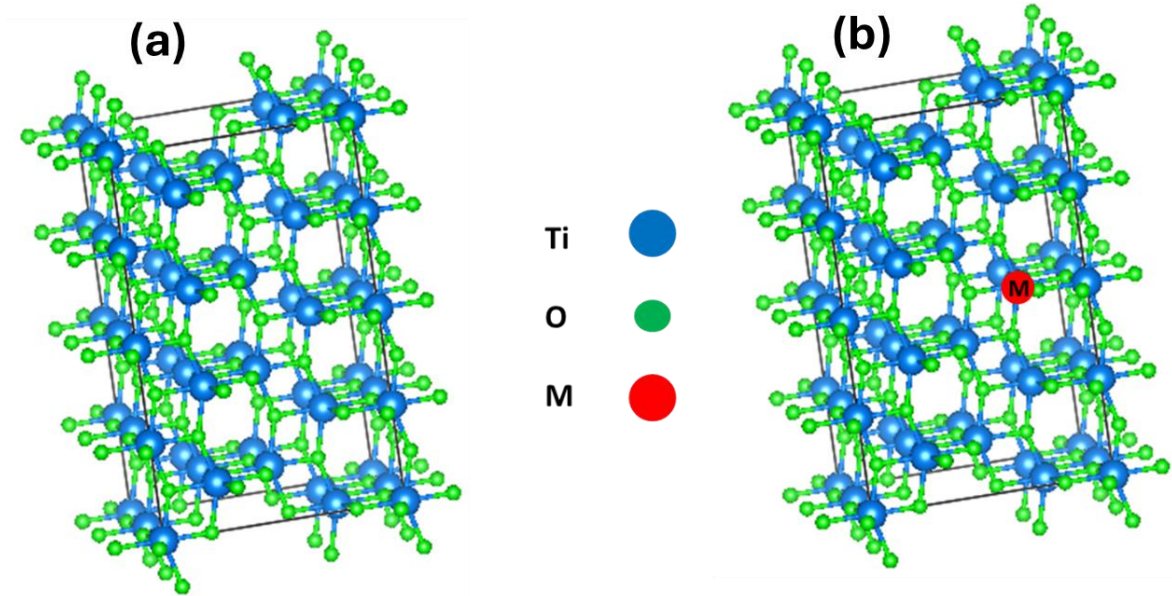

**Figure 1:** *Anatase structure of (a) Pure TiO$_2$ and (b) Doped TiO$_2$ with 2x2x1 supercell. Blue green and red color indicates Ti, O and M (M= Sc, V, Cr, Mn, Fe, Co, Ni, Zn) atoms, respectively.*

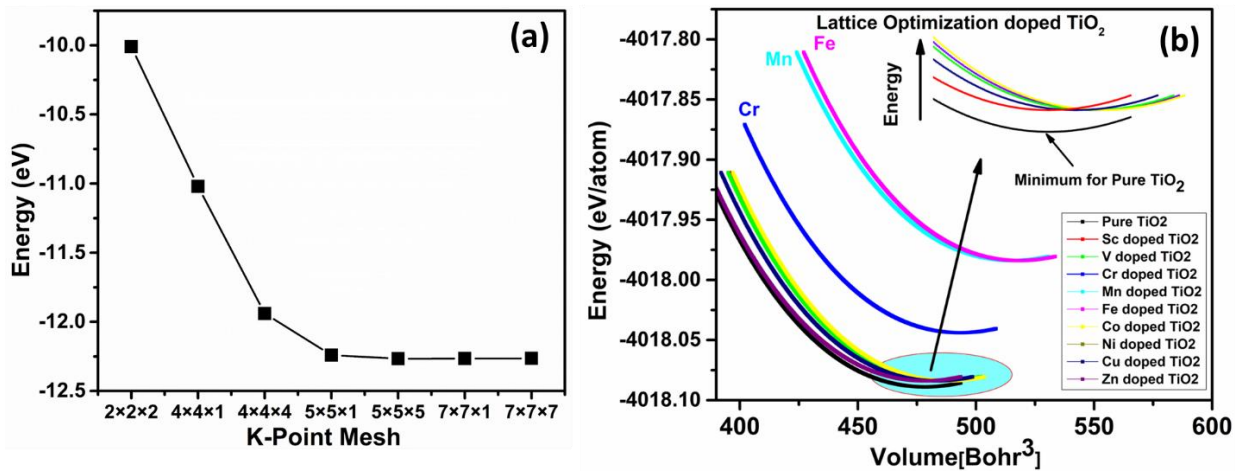

**Figure 2:** *(a) K-point Mesh convergence and (b) Structural optimization of pure and doped TiO$_2$.*

***Table-1:*** *Estimation of Band gap of pure TiO$_2$ with various approximations.*

| Approximations | Pure TiO$_2$ Band Gap (in eV) |
| --- | --- |
| LDA | 2.51 |
| GGA-PBE | 2.72 |
| HSE06 | 3.20 (~ close to experimental value 3.22 eV[4,37]) |

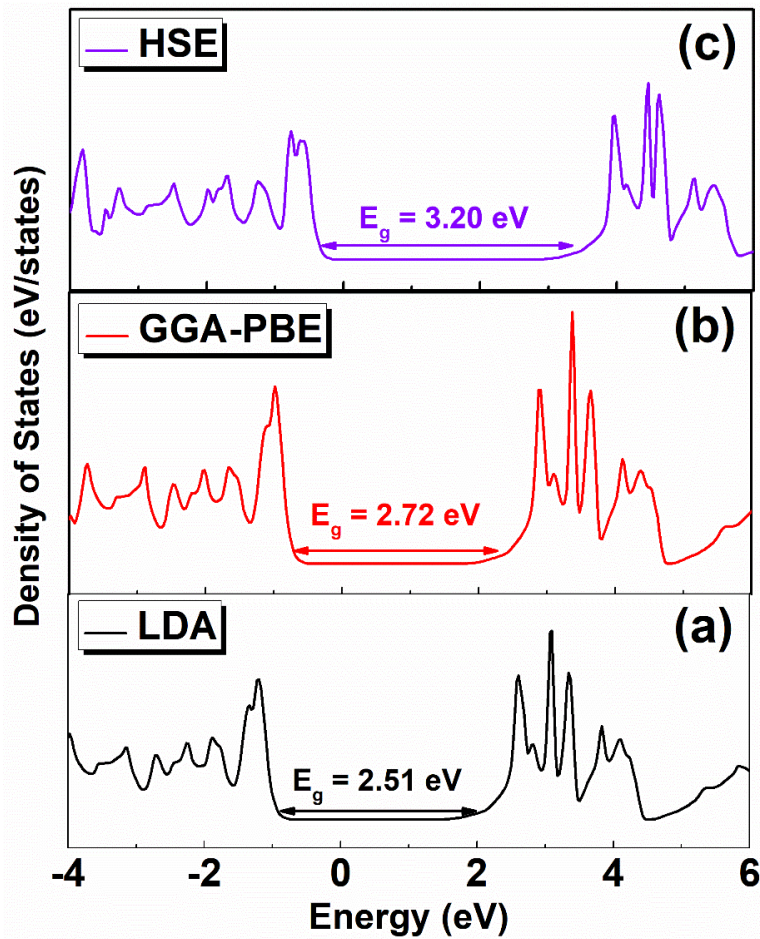

***Figure-3:*** *Total density of states (TDOS) for pure TiO$_2$ using (a) LDA (b) GGA (using PBE functional) and (c) HSE06 approximations.*

In Figure 3 we have plotted calculated TDOS, obtained using various exchange-correlation functionals, for the pure TiO$_2$. In the case of pure TiO$_2$, the energy states near the Fermi level (E$_f$) are primarily composed of the hybridized O-2p and Ti-3d orbitals (see Fig. 4(a)). Previously, it was reported that in the case of pure TiO$_2$, a more detailed examination of the Partial Density of States (PDOS) around the Fermi energy (E$_f$) reveals that the Ti-3d (t$_{2g}$) states exhibit a greater dominance and play a more significant role in the bonding compared to the Ti-3d (e$_g$) states, as determined using the GGA-based calculations[4, 38,39-42]. In the context of O-2p orbitals, it can be observed that the contributions of p$_x$, p$_y$, and p$_z$ orbitals near the Fermi energy (E$_f$) are of comparable magnitudes. The PDOS of pure TiO$_2$, as depicted in Fig. 4(a), has a distinct band gap of 3.20 eV between the VBM and CBM. This value for the band gap (E$_g$) aligns with experimental findings. Based on Fig. 4(a), it is evident that the emergence of a band gap in TiO$_2$ can be attributed to the hybridization of the O-2p (valence band maximum) orbital with the Ti-3d (conduction band minimum) orbital [1,4,43].

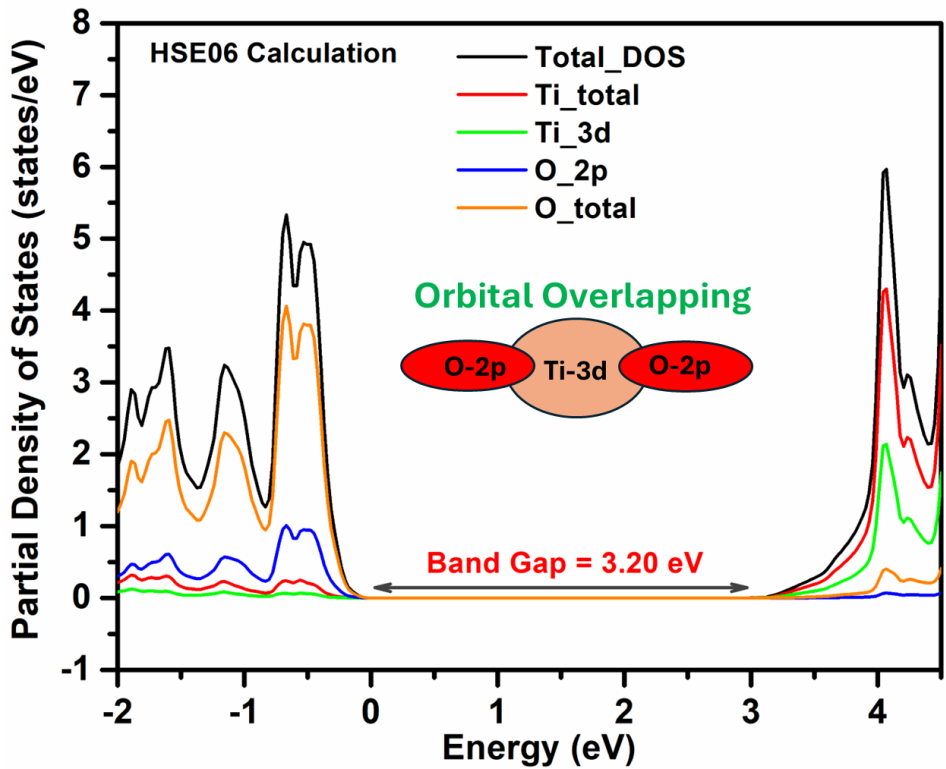

***Figure-4(a):*** *Partial density of states (PDOS) of pure anatase TiO$_2$ shows an electronic band gap of 3.20 eV, with the Ti-3d orbital exhibiting the maximum contribution near the conduction band minimum, while the O-2p orbitals dominate near the valance band maxima. The states near the Fermi energy display hybridization between O-2p and Ti-3d orbitals.*

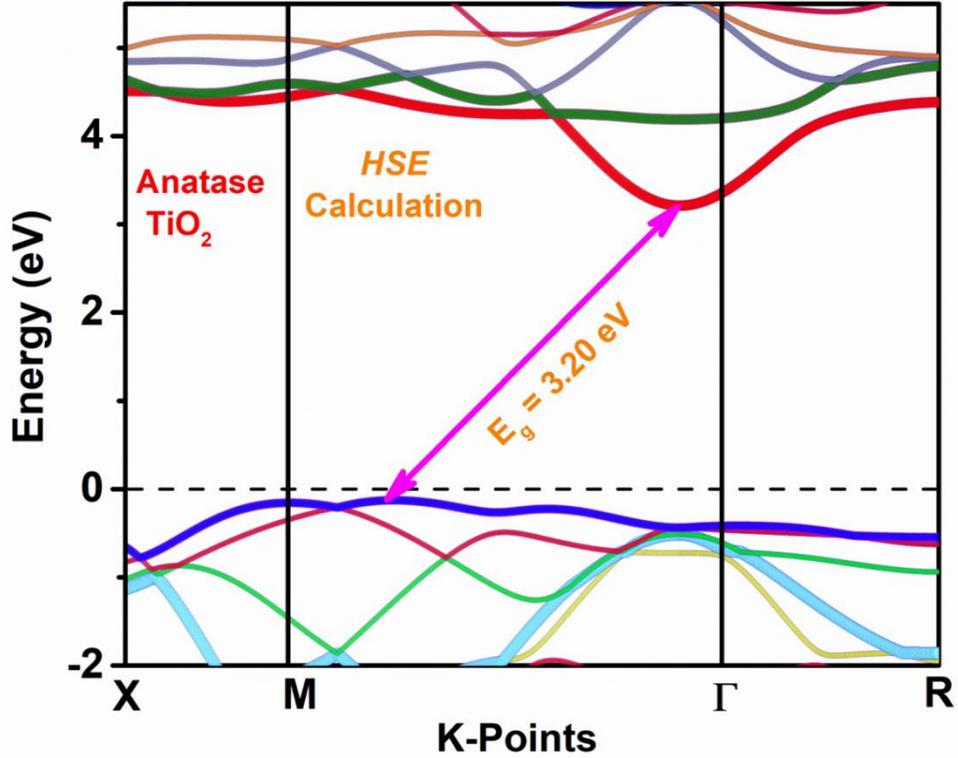

***Figure-4 (b):*** *Band structure of the pure anatase TiO₂ computed using the HSE06 functional. Clearly, the band gap is indirect in nature.*

To further confirm the orbital contribution and the nature of the band gap, band structure calculations for the pure TiO$_2$ have been depicted in Figure 4(b). From Fig. 4(b), It is clearly shown that anatase TiO$_2$ has an indirect nature of band gap of 3.20 eV between Γ and M points. To assess the structural stability of the doped systems, it is important to calculate their formation energies, for which we used the formula [37,38]

$$E_{form} = E_{Doping} - E_{Pure} + \sum_i n_i \mu_i, \qquad (1)$$

where, $E_{form}$ represents the formation energy of the given structure, $E_{Doping}$ represents the total energy of the doped structure, $E_{pure}$ represent the total energy of the pure structure, while n$_i$ and μ$_i$ denote the total number of the i-th atom, and its chemical potential. By including the contributions of the translational, rotational, and vibrational degrees of freedom of a material, it is possible to compute its free energy. However, given the high computational costs of such calculations, we have performed all our calculations at 0K temperature (i.e., avoided the room temperature calculations ~ 298 K) [44-46]. Hence, we have used the reported values of chemical potential [45, 46] needed in equation-1, and with the help of calculated total energy of the pure and doped structures, we have estimated their formation energies.

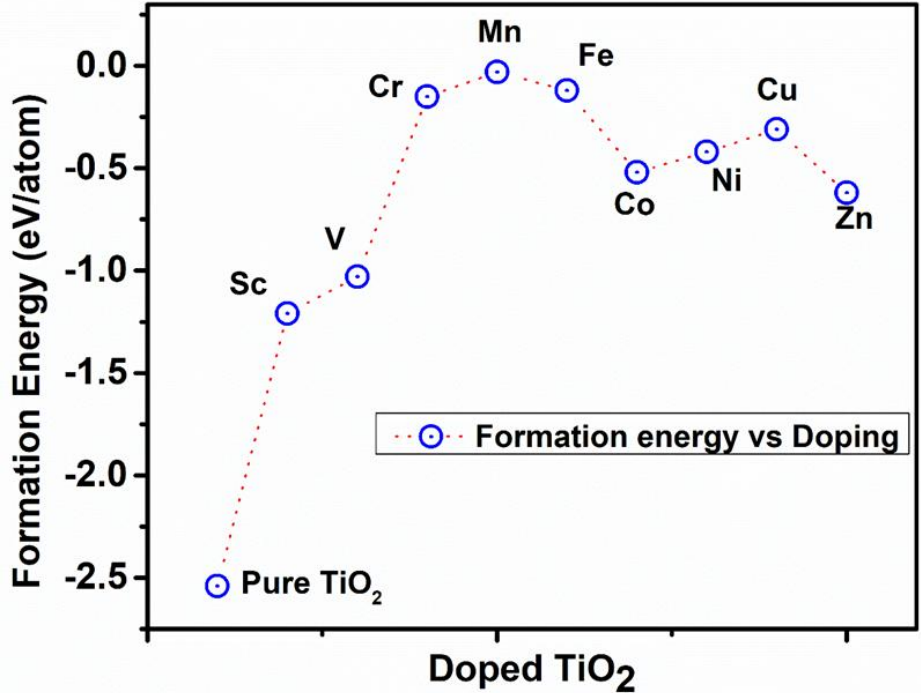

*Figure 5:* *The calculated formation energies of pure and 3d-doped TiO$_2$.*

The calculated formation energies are presented in Fig. 5 from which it is obvious for all the cases the formation energies are negative indicating that it is possible to synthesize these structures in a laboratory [17,43]. The spin-polarized density of states (SPDOS) of the 3d-transition metal doped structures of the anatase TiO$_2$, i.e., Ti$_x$M$_{1-x}$O$_2$, have been computed using the HSE06 functional [4], and presented in Figure 6 (a-i). It is observed that several extra states appear in between valance band and conduction bands for different dopants (i.e. impurity states). Consequently, the number and the nature of the midgap impurity states will be different for each dopant. Furthermore, the occupied and the virtual states available for the eletctronic transitions will also be different as compared to the pure titanium dioxide (TiO$_2$), leading to distinct electronic properties for the 3d transition metal doped TiO$_2$[47,48].

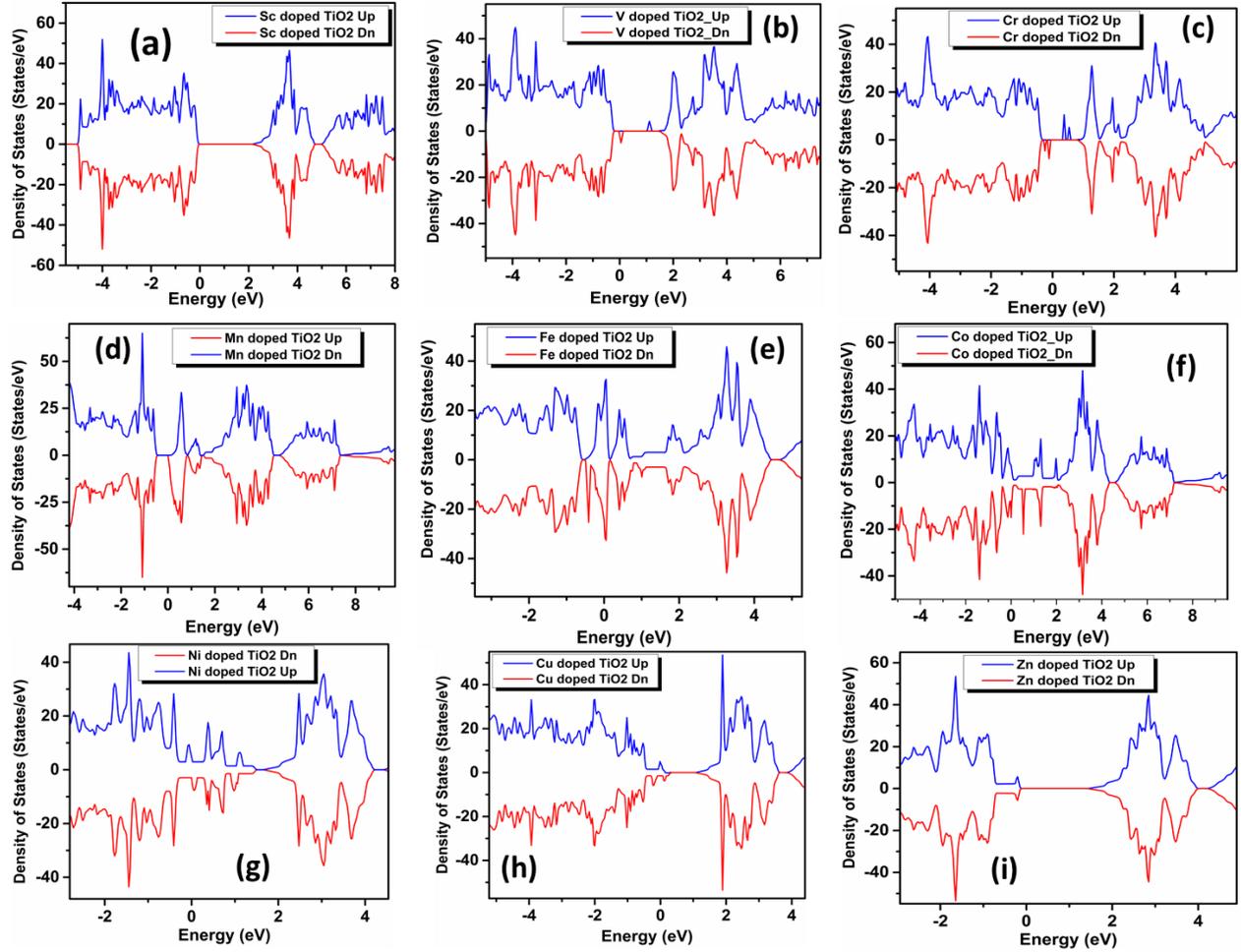

***Figure-6:*** *Spin polarized density of states (SPDOS) for 3d-transition metal doped $TiO_2$ with 2x2x1 supercells: (a) $Ti_{15}Sc_1O_{32}$ (b) $Ti_{15}V_1O_{32}$ (c) $Ti_{15}Cr_1O_{32}$ (d) $Ti_{15}Mn_1O_{32}$ (e) $Ti_{15}Fe_1O_{32}$ (f) $Ti_{15}Co_1O_{32}$ (g) $Ti_{15}Ni_1O_{32}$ (h) $Ti_{15}Cu_1O_{32}$ (i) $Ti_{15}Zn_1O_{32}$.*

In the SPDOS, the shifting of the electronic states from CBM and VBM to the mid-gap region with the increase in the atomic number of the the dopant atoms clearly implies the existence of variable oxidation states of 3d-transition metal atoms (Sc=+3, V= +2,+3,+4,+5, Cr= +2,+3,+4,+6, Mn= +2,+3,+4,+6,+7 Fe=+2,+3+6, Co=+2,+3, Ni=+2,+3, Cu= =1,+2,+3, Zn=+2) causing the hybridization between O-2p and Ti-3d to change. Different dopants (i.e. transition metal atoms) have different number of unpaired electrons in their 3d orbital and hence we obtain different up- and down- SPDOS, leading to the onset of ferromagnetic behavior[9,19,49]. Band structure calculations for three representative cases (a) V (b) Mn, and (c) Ni doped $TiO_2$ have been performed using the HSE06 functional to further confirm the presence of impurity states in between CBM and VBM, as shown in figure 7 (a)-(c). From the band structure it is obvious that the total number of mid-gap states increase from one to three as the dopant changes from V to Mn, pointing to their dependence on the number of dopant 3d electrons.

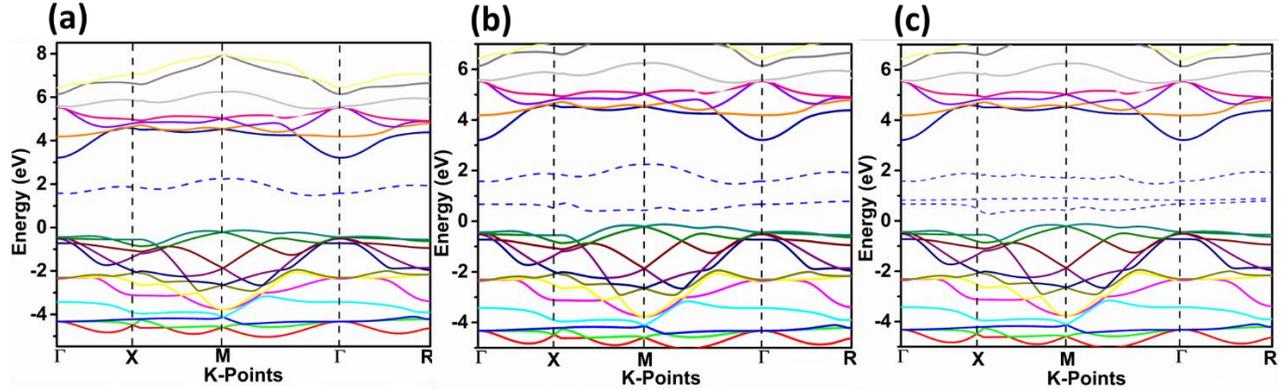

*Figure 7:* Band structure of (a) V (b) Mn and (c) Ni doped $TiO_2$ computed using the HSE06 functional. Solid lines represent electronic states corresponding to bulk bands, while the dashed lines represent defect states due to the corresponding dopants.

We have also calculated magnetic moments for all cases of doping, and the results are presented in Table 2, and we find that the maximum magnetic moments of 2.45$\mu_B$, 3.57$\mu_B$, 4.60$\mu_B$, respectively, are induced at the doping sites for the dopants Cr, Mn and Fe. This clearly is due to a large number of unpaired electrons in 3d orbital of these atoms, consistent with the Hund's rule. Due to orbital hybridization, the neighboring oxygen atom also acquires very small magnetic moments.

*Table-2* Calculated magnetic moments of metal doped $TiO_2$ at: (a) dopant sites, and (b) O atoms nearest to the dopant atoms

| System | (a) Magnetic moment at the dopant site (in $\mu_B$) | (b) Magnetic moment at the O atom nearest to the dopant atom (in $\mu_B$) |
|---|---|---|
| $Ti_{15}Sc_1O_{32}$ | 0 | 0($O_1$),0($O_2$) |
| $Ti_{15}V_1O_{32}$ | 0.9 | 0.01($O_1$),0.01($O_2$) |
| $Ti_{15}Cr_1O_{32}$ | 2.45 | 0.02($O_1$),0.03($O_2$) |
| $Ti_{15}Mn_1O_{32}$ | 3.57 | 0.02($O_1$),0.03($O_2$) |
| $Ti_{15}Fe_1O_{32}$ | 4.60 | 0.04($O_1$),0.04($O_2$) |
| $Ti_{15}Co_1O_{32}$ | 1.42 | 0.05($O_1$),0.04($O_2$) |
| $Ti_{15}Ni_1O_{32}$ | 0.10 | 0.02($O_1$),0.03($O_2$) |
| $Ti_{15}Cu_1O_{32}$ | 1.02 | 0.02($O_1$),0.03($O_2$) |
| $Ti_{15}Zn_1O_{32}$ | 0.34 | 0.01($O_1$),0.01($O_2$) |

## Conclusions

In conclusion, to see the effect of doping of transition metal ions in anatase $TiO_2$, a comparative study has been performed using a density functional approach. The electronic structure of the 3d-transition metal doped $Ti_xM_{1-x}O_2$ (M=Sc, V Cr, Mn, Fe, Co, Ni, Cu, Zn) has been examined, and values of band gaps estimated using both the PBE functional within the GGA scheme, and the hybrid HSE06 functional. The presence of midgap states is a consequence of doping, and for the case of doping with Cr, Mn, and Fe, our calculations predict large magnetic moments at the dopant sites. Therefore, we believe that Cr-, Mn-, and Fe-doped anatase $TiO_2$ can serve as effective dilute magnetic semiconductors, with tremendous possibilities of spintronic applications.

## Acknowledgments

VM sincerely thank to the Department of Physics, MIT Manipal for providing funding from the institute. Indian Institute of Technology Bombay, India is acknowledged by one of the authors (SP) for providing the Institute Postdoctoral fellowship.

## Compliance with ethical standards:

Conflict of interest: The authors declare that they do not have any conflict of interest.

## Data availability statement –

The raw/processed data required to reproduce these findings cannot be shared at this time as the data also forms part of an ongoing study.